\documentstyle[aps,prl,twocolumn,psfig]{revtex}

\begin{document}

\newcommand{\tit}
{Kaon Condensation and Dynamical Nucleons\\ in Neutron stars}
\newcommand{\autha} {Norman K. Glendenning}
\newcommand{\authb} {J\"urgen Schaffner-Bielich}
\newcommand{\authc} {F. Weber}
\newcommand{\lbl}{LBNL-42101}
\newcommand{\dateofdoc}{September 2, 1998}
\newcommand{\doe}
{This work was supported by the
Director, Office of Energy Research,
Office of High Energy
and Nuclear Physics,
Division of Nuclear Physics,
of the U.S. Department of Energy under Contract
DE-AC03-76SF00098.}

\begin{titlepage}
\parbox{3.5in}{\begin{flushleft}Physical Review Letters (to be published)
\end{flushleft}}%
\parbox{3.5in}{\begin{flushright} \lbl \end{flushright}}
~\\[3ex]
\begin{center}
\begin{LARGE}
\renewcommand{\thefootnote}{\fnsymbol{footnote}}
\setcounter{footnote}{1}
\tit {\footnote{\doe}}\\[3ex]
\end{LARGE}

\renewcommand{\thefootnote}{\fnsymbol{footnote}}
\setcounter{footnote}{2}
\begin{Large}
\autha~~and~~ \authb \\[2ex]
\end{Large}
\dateofdoc \\[2ex]
\end{center}

\begin{center}
{\Large
Nuclear Science Division\\ and \\Institute for Nuclear \& 
 Particle Astrophysics\\ Lawrence Berkeley National Laboratory\\[2ex]
  Berkeley, California}\\[10ex]
  \begin{figure}[htb]
  \begin{center}
  \leavevmode
  \hspace{-.2in}
  \psfig{figure=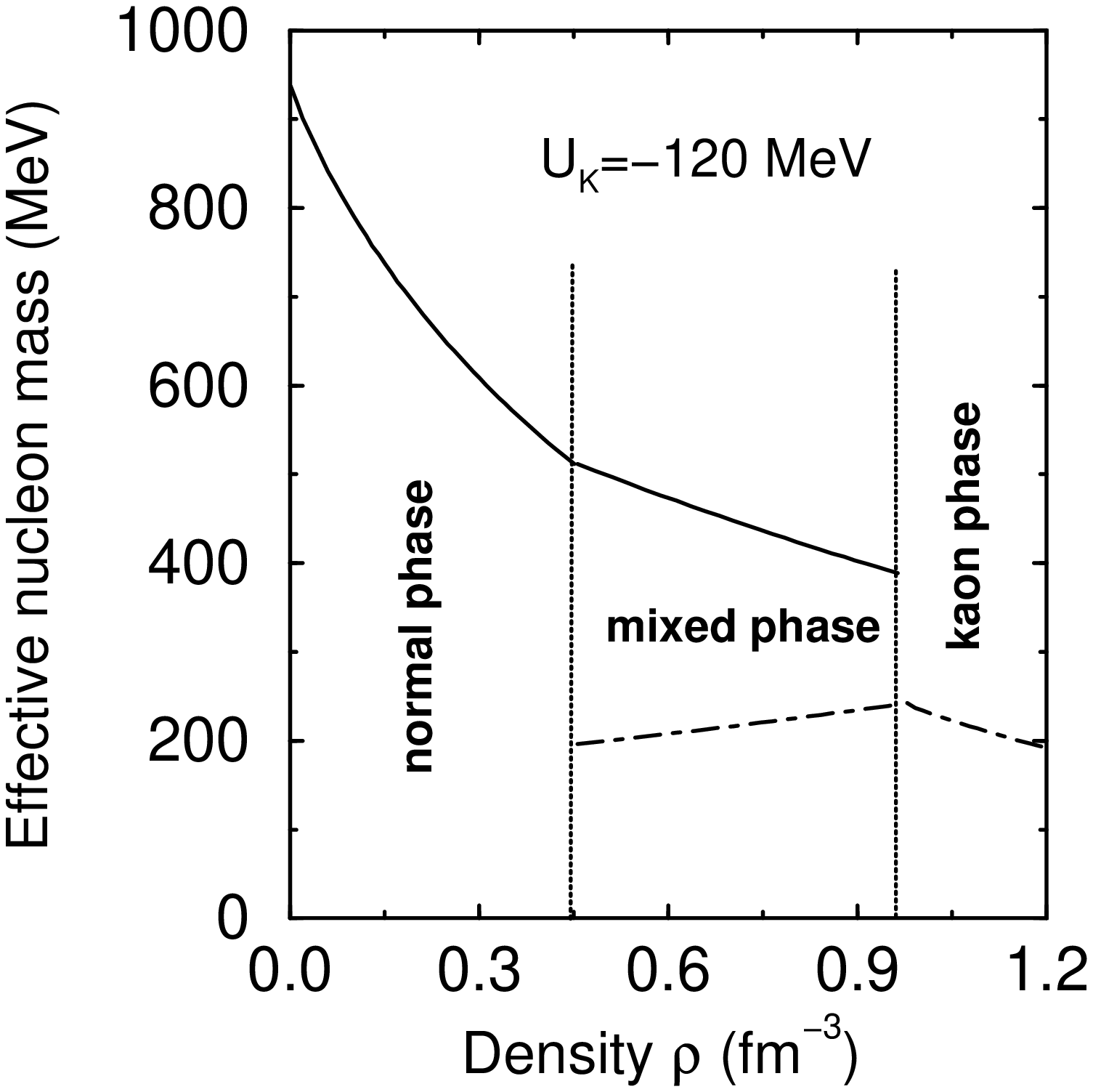,height=3.2in}
  \end{center}
  \end{figure}

   \end{center}
    \end{titlepage}
    \clearpage
    \eject
    \newpage

\tightenlines
\title{Kaon Condensation and Dynamical Nucleons\\in Neutron stars}
\author{Norman K. Glendenning and J\"urgen Schaffner-Bielich}
\address{
Nuclear Science Division and
Institute for Nuclear \& Particle Astrophysics,
Lawrence Berkeley National Laboratory,\\
University of California, Berkeley, CA 94720, USA}
\date{September 2, 1998}
\maketitle

\begin{abstract}
We discuss the nature of the kaon condensation phase transition. We find
several remarkable features which, if kaons condense in neutron stars, will
surely effect superfluidity and transport properties. The mixed phase region
occupying a large fraction of the star is
permeated with microscopic drops located at lattice sites of one phase immersed
in the background of the other phase. The electric charge in the drops is
opposite to that of the background phase {\sl and} nucleons have a mass
approximately a factor two different depending on whether they are in the drops
or the background phase.
\end{abstract}
\draft

Kaon condensation as well as pion condensation have been the focus
of a number of papers over the years 
\cite{Migdal78,hartle75:a,weise75:a,migdal79:a,migdal79:b%
 ,weise79:a,glen85:b,KN86,Brown94,Lee96}. 
The possible presence of a 
kaon condensate in neutron stars
has received special attention recently
\cite{Brown92,Thorsson94,Fujii96,Li97}. 
However two important
facets of the problem have not been realized---the dynamical
nature of the
nucleons, as well as stability with respect to the electric charge
at the interface between
phases.
Both facets and their structural consequences for the
true nature of the kaon condensed phase are elucidated here.

The effect of the nuclear medium on the energy of a test kaon 
has been studied by a number of authors and 
it is believed that the interactions
reduce the K$^-$ energy as a function of baryon density
\cite{kampfer95:a,Koch94,Waas97}. Kaonic data support the
conclusion that there is a highly
attractive kaon optical potential in dense nuclear matter \cite{Fried94}. 
If the
kaon energy intersects the electron chemical potential at some density, kaons
thereafter will be energetically more favorable than electrons
as the neutralizing
agent of positive charge \cite{glen85:b}.
The very interaction that reduces
the kaon energy modifies the nucleons with which they interact.
Therefore, the kaon condensed phase is not merely nuclear matter in which
kaons are present---it is a fundamentally different phase in every
way, including the masses of nucleons.

Moreover, first order kaon condensation requires careful attention
to the manner in which phase equilibrium is found so 
as to ensure a stable configuration. The Maxwell construction
is inadequate for this purpose in neutron star matter, which has
two conserved charges---baryon number and electric charge (which vanishes
globally).
 The Maxwell construction, which has formerly been
employed in treatments of both
first order pion
\cite{hartle75:a,weise75:a,migdal79:a,migdal79:b,weise79:a}, 
and kaon condensation
\cite{Brown92,Thorsson94,Fujii96,Li97}, 
can insure the equality of only one chemical potential in the two phases
in equilibrium \cite{glen91:d,glen91:a}.
Therefore, if the equality of the baryon chemical potential has been 
insured there must be a potential difference in the 
electron chemical potential at the interface of the two phases.

Details in the matters discussed in this paper will of course depend
on the particular nuclear model and the kaon interaction with nucleons.
But the physics will be similar for any physically correct implementation.
So as to emphasize the impact of kaon condensation we neglect the possible
suppression by hyperons 
\cite{glen85:b,schaffner96:a,prakash95:a,knorren95b}.
For the nucleon sector,
we choose the standard nuclear field theory of nucleons
interacting through scalar, vector and vector iso-vector
mesons ($\sigma, \omega {~\rm and~} \rho$), solved in the mean field
approximation \cite{book}. 
The kaon is then coupled to the vector
meson fields using minimal coupling
\begin{equation}
{\cal L}_K = {\cal D}_\mu^* K^* {\cal D}^\mu K - {m^*_K}^2 K^*K
\label{eq:lagK}
\end{equation}
where 
\begin{equation}
{\cal D}_\mu = \partial_\mu + i g_{\omega K} V_\mu +
i g_{\rho K} \vec{\tau}_K \vec{R}_\mu
\quad .
\end{equation}
The kaon and scalar fields are coupled through
\begin{equation}
m^*_K = m_K - g_{\sigma K} \sigma \quad .
\label{eq:scoupl}
\end{equation}
We choose to couple the vector fields according to the simple quark
and isospin counting rule
\begin{equation}
g_{\omega K} = \frac{1}{3} g_{\omega N} \quad \mbox{ and } \quad
g_{\rho K} = g_{\rho N}
\quad.
\end{equation}
The scalar coupling constant
$g_{\sigma K}$ is fixed to the optical potential of the K$^-$ at
normal nuclear density
$\rho_0$:
\begin{equation}
U_K (\rho_0) = -g_{\sigma K} \sigma( \rho_0) - g_{\omega K} V_0 (\rho_0)
\quad . \label{optical}
\end{equation}
Coupled channel calculations find an optical potential between $-100$ MeV 
\cite{Koch94} and $-120$ MeV \cite{Waas97}. We choose the latter value in
the following. 
The five coupling constants in the nucleon sector are determined
algebraically \cite{book}
to reproduce the saturation properties of symmetric
nuclear matter;
$E/A=-16.3$ MeV, $\rho_0=0.153$
fm$^{-3}$, $a_{{\rm sym}}= 32.5$ MeV, $K=240$ MeV, and $m^*/m=0.78$. 

The equation of motion for the kaon can be written as
\begin{equation}
\left[{\cal D}_{\mu} {\cal D}^\mu
 + {m^*_K}^2 \right] K = 0
 \quad .
 \end{equation}
 One gets then a dispersion relation for the K$^-$
 for s-wave condensation ($\vec{k}=0$) of the form
 \begin{equation}
 \omega_K = m_K - g_{\sigma K}\sigma -
 g_{\omega K} V_0 - g_{\rho K} R_{0,0} 
 \quad .
 \label{eq:disp}
 \end{equation}

 \begin{figure}[b]
 \vspace{-.4in}
 \begin{center}
 \psfig{figure=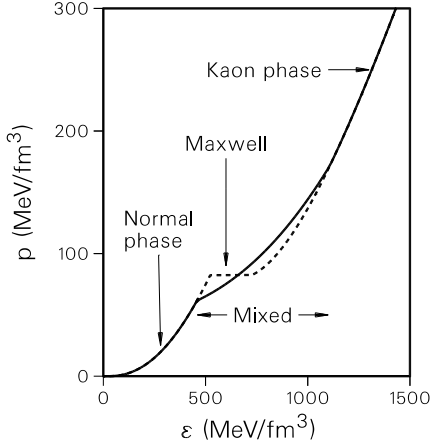,height=2.5in}
 \vspace{.3in}
 \caption { \label{EOS_KAON_120}
 The equation of state with the three phases indicated.
 Pressure increases  monotonically  for Gibbs equilibrium.
 The pressure range of the mixed phase exceeds that of the
 normal phase. The Maxwell construction is shown for comparison.}
 \end{center}
 \vspace{-.25in}
 \end{figure}

\begin{figure}[b]
 \vspace{-.4in}
 \begin{center}
 \psfig{figure=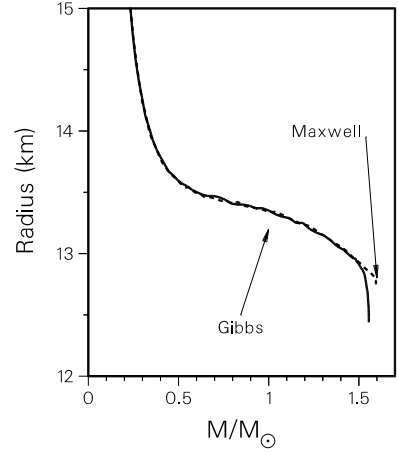,height=2.5in}
 \vspace{.3in}
 \caption { \label{RM_KAON_120} Mass-radius relation of neutron stars
  corresponding to
 Fig.\ \ref{EOS_KAON_120}.}
 \end{center}
 \vspace{-.25in}
 \end{figure}

It is possible to satisfy 
 Gibbs conditions for phase equilibrium simultaneously with conservation
 laws
 for substances of more than one conserved charge {\sl only}
 by applying the
 conservation law(s) in a {\sl global} rather than a {\sl local}
 sense \cite{glen91:d,glen91:a}.
 Thus for neutron star matter, which has two conserved
 charges, the Gibbs conditions and conservation of electric charge
 read
 \begin{eqnarray}
 p_N(\mu_B, \mu_e) =p_K(\mu_B, \mu_e) \label{gibbs} \\
 q_{{\rm total}}= (1-\chi)q_N(\mu_B, \mu_e) +\chi q_K(\mu_B, \mu_e) = 0\,.
 \label{conservation}
 \end{eqnarray}
 where $q_N$ and $q_K$ denote  the charge densities including leptons
 in the normal and kaon condensed phase,
 $\mu_B$ the baryochemical potential and
 $\mu_e$ the electrochemical potential.
 This pair of equations can be solved for $\mu_B$ and $\mu_e$
 for any proportion of kaon phase $\chi$ in the interval (0,1).
 Therefore the chemical potentials are functions of proportion $\chi$
 and therefore also  all other properties of the two phases, including,
 very importantly,
 the common pressure.
 The total baryon density for the chosen $\chi$
 and corresponding chemical potentials is given by
 \begin{equation}
 \rho_{{\rm total}}= (1-\chi)\rho_N(\mu_B, \mu_e) +\chi \rho_K(\mu_B, \mu_e)
 \label{density}
 \,,
 \end{equation}
 where $\rho$ denotes baryon density.
 The equation of state is shown in Fig.\ \ref{EOS_KAON_120}
 and compared with that of a Maxwell construction and the corresponding
  sequence of neutron stars in  Fig.\ \ref{RM_KAON_120} where the maximum mass
  can be read as $1.56 M_\odot $.
 Since the pressure 
 varies monotonically in a star, a mixed phase of the Maxwell type would
 be totally absent  because the constant pressure is mapped
 onto a single radial point. Gravity squeezes out a constant
 pressure mixed phase. This has been so in all previous
 treatments of first order kaon condensation \cite{Thorsson94,Fujii96,Li97}
as well as pion condensation
\cite{hartle75:a,weise75:a,migdal79:a,migdal79:b,weise79:a}.

\begin{figure}[htbp]
{\hspace*{-0.4in}\psfig{figure=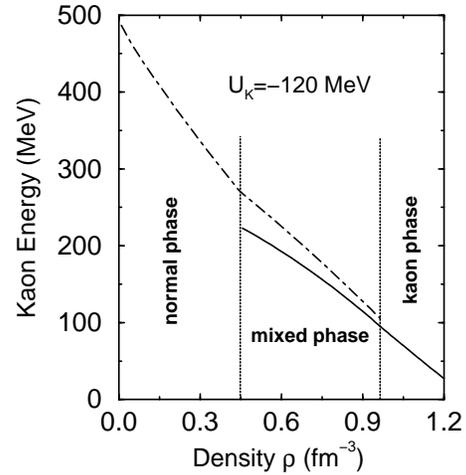,height=0.29\textheight}}
\caption{\protect\label{kaon_energy}
Dashed line  shows the medium modified  energy of a test kaon.
The energy of the real
kaons begins in the condensed fraction of the mixed phase (solid line).}
\end{figure}

We can trace the density dependence of  energy  $\omega_K$ of a
test kaon in the normal medium as depicted in Fig.\ \ref{kaon_energy}.
The kaon energy 
decreases with increasing density because of the attractive
vector potential. At some density it
may intersect the electron chemical potential.
At the density for which the
equality 
 $\omega_K=\mu_e$
 first holds, kaons of modified mass occupy a small volume fraction
of the total medium with a number density of
\begin{eqnarray}
n_K=
2\left(\omega_K + g_{\omega K} V_0 + g_{\omega K} R_{0,0}\right)
K^* K = 2 m^*_K K^*K \quad .
\end{eqnarray}
The energy of these medium modified kaons is less
than that of a test
kaon in the normal phase even when the two phases are in equilibrium,
as seen in Fig.\ \ref{kaon_energy}.

\begin{figure}[htbp]
{\hspace*{-0.4in}\psfig{figure=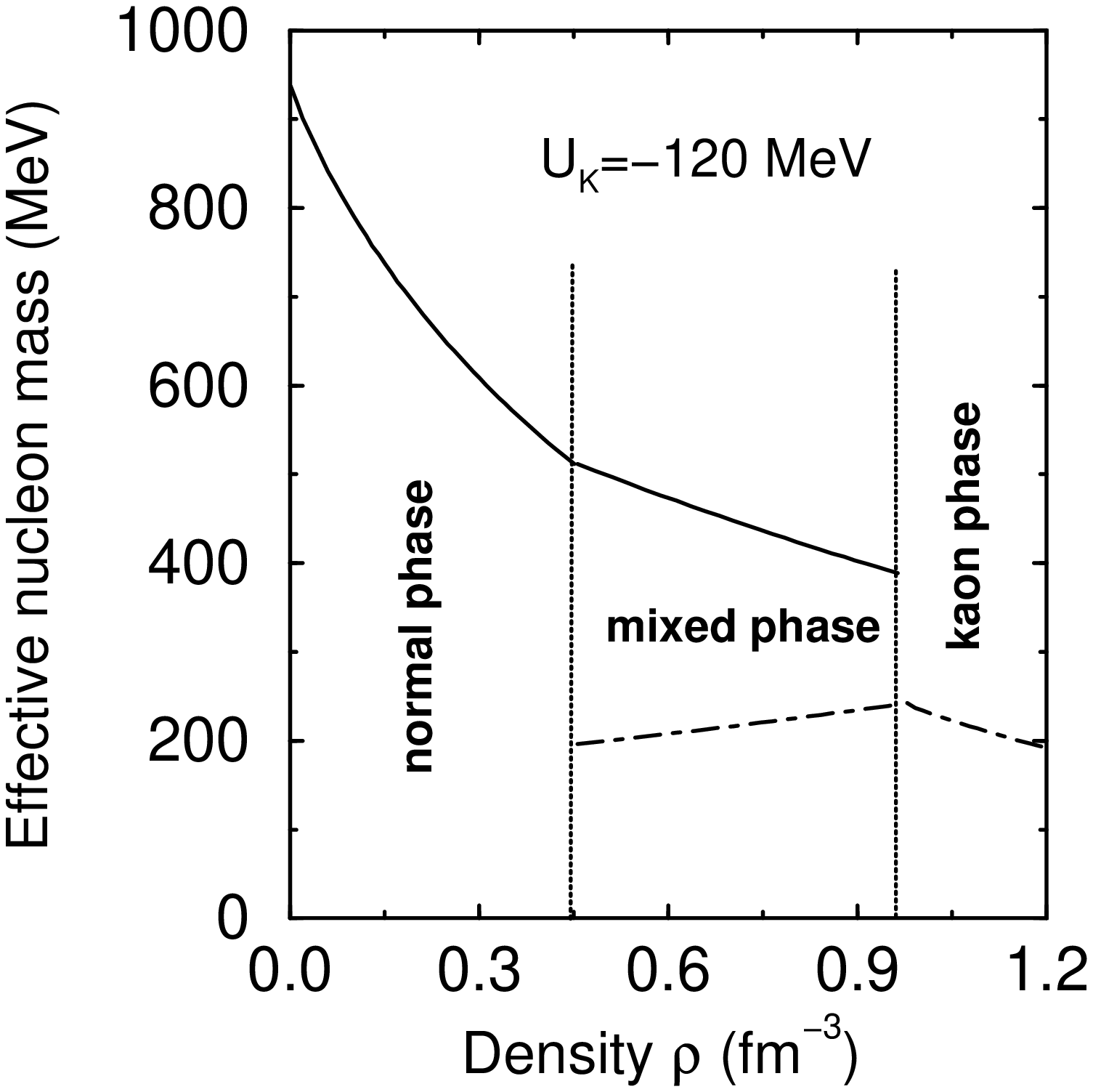,height=0.29\textheight}}
\caption{\protect\label{effmass}
The effective nucleon mass in the three phases.
The nucleon masses differ by a factor of two in the regions of the
mixed phase occupied by normal
and kaon  phases respectively, being smaller in the latter.}
\end{figure}

The kaon-nucleon interaction
effects not only the kaon mass in the medium, but also the nucleon mass.
Therefore, in the mixed phase of normal and condensed
matter, the nucleon mass is different by a factor of more than
two, as seen in Fig.\ \ref{effmass},
depending on which region of the mixed phase the 
nucleon is in.

The imposition of charge conservation as a global constraint,
as in (\ref{conservation}), rather than as a local constraint, is not 
only necessary
so that Gibbs equilibrium can be achieved---it is also
necessary so as to find the lowest energy state \cite{glen91:d,glen91:a}.
In general, the lowest energy state in the mixed phase of a substance
of more than one conserved charge is achieved by exchange of conserved
quantities between the phases in equilibrium, the exchange being driven by 
an internal force. In the case of neutron star matter, the internal
force is the isospin driving force to which both the
Fermi energies and the specific nuclear force (coupling of the rho
meson to the isospin) contribute.
The normal phase of neutron star matter is highly isospin asymmetric
because of the constraint of charge neutrality imposed by gravity.
The isospin symmetry
energy of nuclear matter drives a redistribution of charge between the
 normal and condensed phase as soon as some of the latter
 is formed. In this way the repulsive isospin energy can be reduced
 and the normal phase becomes more
 symmetric in neutrons and protons
as the proportion of kaon phase  increases.

\begin{figure}[htb]
\vspace{-.4in}
\begin{center}
\psfig{figure=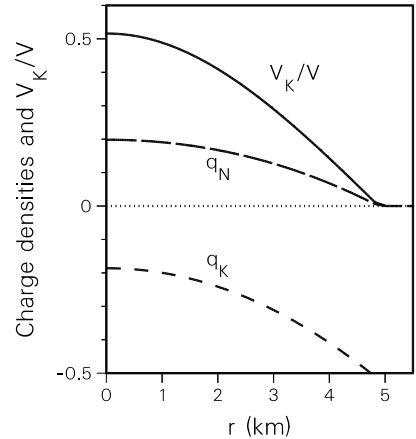,height=2.5in}
\vspace{.3in}
\caption{ \label{CHI_KAON_120} Charge density on regions
of normal and kaon condensed phase in the region of mixed phase in 
a neutron star at the mass limit. The volume weighted sum is
zero. The fraction of the kaon phase is also shown.}
\end{center}
 \vspace{-.25in}
\end{figure}

 For the above reason, 
 the kaon phase is maximally negatively charged when its volume fraction 
 is small
 and the charge density of this phase approaches zero as it 
 becomes the dominant phase. 
 Such behavior is easy to understand: under the constraint
 of charge neutrality, the total energy
 can be optimized by a configuration that places 
 the dominant phase closest to its
 lowest energy, even  at the expense of the rarer phase being far from
 its optimum.
 The volume weighted charge densities sum to zero as
  required for a stellar medium which must be neutral on account
 of the disproportionate strength of the long-range forces---Coulomb
 and gravitational. The oppositely charged
 densities as a function of position
 in the star
 are shown in Fig.\ \ref{CHI_KAON_120} and are seen to vary strongly
 as a function of position in the star as the kaon phase becomes
 more dominant at greater depth.
 Of course, the charge density vanishes identically
 in the pure homogeneous phases.

\begin{figure}[htb]
 \vspace{-.4in}
\begin{center}
\psfig{figure=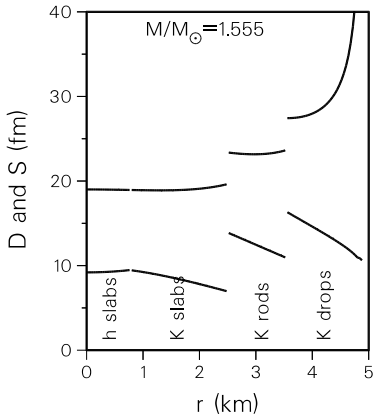,height=2.3in}
 \vspace{.3in}
 \caption{ \label{crys_kaon_120} The diameter (D)
 and spacing (S) of geometrical
 structures at lattice sites in the mixed phase are shown
 as a function of position in the star. Shapes of the
 rare phase object immersed in the dominant phase background are denoted by
 K drops for kaon condensed phase droplets, and h slabs for normal hadronic
 phase slabs, etc.}
 \end{center}
  \vspace{-.25in}
  \end{figure}

The mixed phase of normal and kaon condensed matter, having opposite
electric charge, will form a
Coulomb lattice of 
the rarer phase 
immersed in the dominant. The competition between Coulomb and surface
interface energy will determine the size of the rare phase objects,
their geometry and their spacing in the background of the dominant phase
\cite{glen91:d,glen91:a,heiselberg93:a,glen95:c,glen97:c}. 
The idealized
geometrical structures that appear in our model star
are illustrated in Fig.\ \ref{crys_kaon_120}, which shows the diameter (D)
of objects at lattice sites
and their spacing (S) as a function of position in the star.
 Sizes and spacings depend on the
 surface tension as $\sigma^{1/3} $ and is proportional to the difference in 
 the energy densities of the phases in contact
 \cite{glen95:c}. 
For the present choice of parameters the
pure kaon phase does not appear and the geometrical phases terminate 
with slabs of normal hadronic matter embedded in the kaon condensed phase. 
If the optical potential (\ref{optical}) were stronger, the geometries
would span the  full range from kaon-condensed-phase drops
immersed in normal matter to the inverse,
and the pure kaon condensed phase would occupy 
the core of the star.
Of course, further investigations are needed to disclose details of the
geometrical structures, including, for example a computation of the
surface tension between phases.

From the above results
the following picture
emerges.
If kaons condense in neutron stars, there will exist
a region where normal matter will
coexist in phase equilibrium with the condensed phase. 
The properties of the mixed phase  are
particularly noteworthy.
It will occupy a wide radial extent of the
star because of the wide range of pressure that it exists under.
It is dotted by electrically charged
blobs of one phase 
immersed in the 
other phase of opposite charge, all arranged on a Coulomb
lattice. The size of the objects at the lattice sites,
their geometry and their spacing, will vary with depth in the
star because of the pressure variation.
More than this, the eigenmass of nucleons
are different by a factor two or more
in the background phase than in the blobs
occupying the lattice sites
{\sl and} 
also vary with depth
in the star. All of these remarkable
properties certainly give food for thought as
to what the superfluid and transport properties might be for such
an
unusual configuration of
matter, as well as for the cooling and glitch behavior of pulsars.
The kaon condensation phase
transition might be signaled in pulsar timing just as for
 the deconfined phase transition \cite{glen97:a}.

{\bf Acknowledgments:}
J. S.-B. acknowledges support by the Alexander-von-Humboldt Stiftung. 
This work was supported by the
Director, Office of Energy Research,
Office of High Energy
and Nuclear Physics,
Division of Nuclear Physics,
of the U.S. Department of Energy under Contract
DE-AC03-76SF00098.


\vspace*{-0.2in}
\begin{thebibliography}{10}
\vspace*{-0.5in}

\bibitem{Migdal78}
A.~B. Migdal, Rev. Mod. Phys. {\bf 50},  107  (1978).

\bibitem{hartle75:a}
J. B. Hartle, R. F. Sawyer and D. J. Scalapino, Astrophys.\ J.\ {\bf 199}
  (1975) 471.

\bibitem{weise75:a}
W. Weise and G. E. Brown, Phys.\ Lett.\ {\bf 58B} (1975) 300.

\bibitem{migdal79:a}
A. B. Migdal, A. I. Chernoustan and I. N. Mishustin, Phys.\ Lett.\ {\bf 83B}
  (1979) 158.

\bibitem{migdal79:b}
A. B. Migdal, in {\sl Mesons in Nuclei, Vol III} edited by M. Rho and D.
  Wilkinson (North-Holland Publishing Co., 1979) p. 941, see esp. p. 978.

\bibitem{weise79:a}
S.-O. Backman and W.
Weise in {\sl Mesons in Nuclei, Vol III} edited by M. Rho and
  D. Wilkinson (North-Holland Publishing Co., 1979) p. 1095, see esp. p. 1116.

\bibitem{glen85:b}
N. K. Glendenning, Astrophys.\ J.\ {\bf 293} (1985) 470.

\bibitem{KN86}
D.~B. Kaplan and A.~E. Nelson, Phys. Lett. B {\bf 175},  57  (1986), ibid B
  179, 409 (E).

\bibitem{Brown94}
G.~E. Brown, C.-H. Lee, M. Rho, and V. Thorsson, Nucl. Phys. A {\bf 567},  937
  (1994).

\bibitem{Lee96}
C.-H. Lee, Phys. Rep. {\bf 275},  255  (1996).

\bibitem{Brown92}
G.~E. Brown, K. Kubodera, M. Rho, and V. Thorsson, Phys. Lett. B {\bf 291},
  355  (1992).

\bibitem{Thorsson94}
V. Thorsson, M. Prakash, and J.~M. Lattimer, Nucl. Phys. A {\bf 572},  693
  (1994).

\bibitem{Fujii96}
H. Fujii, T. Maruyama, T. Muto, and T. Tatsumi, Nucl. Phys. A {\bf 597},  645
  (1996).

\bibitem{Li97}
G.~Q. Li, C.-H. Lee, and G.~E. Brown, Phys. Rev. Lett. {\bf 79},  5214  (1997).

\bibitem{kampfer95:a}
E. E. Kolomeitsev, D. N. Voskresensky and B. K\"ampfer, 
Nucl.\ Phys.\ A {\bf 588}
  (1995) 889.

\bibitem{Koch94}
V. Koch, Phys. Lett. B {\bf 337},  7  (1994).

\bibitem{Waas97}
T. Waas and W. Weise, Nucl. Phys. A {\bf 625},  287  (1997).

\bibitem{Fried94}
E. Friedmann, A. Gal, and C.~J. Batty, Nucl. Phys. A {\bf 579},  518  (1994).


\bibitem{glen91:d}
N. K. Glendenning, Nucl. Phys. B (Proc. Suppl.) {\bf 24B} (1991) 110.

\bibitem{glen91:a}
N. K. Glendenning, Phys. Rev. D {\bf 46} (1992) 1274.

\bibitem{schaffner96:a}
J. Schaffner and I. N. Mishustin, Phys.\ Rev.\ C {\bf 53} (1996) 1416.

\bibitem{prakash95:a}
P. J. Ellis, R. Knorren and M. Prakash, Phys.\ Lett.\ {\bf B349} (1995) 11.

\bibitem{knorren95b}
R. Knorren, M. Prakash, and P.~J. Ellis, Phys. Rev. C {\bf 52},  3470  (1995).

\bibitem{book}
N. K. Glendenning, {\sl COMPACT STARS, Nuclear Physics, Particle Physics, and
  General Relativity} (Springer--Verlag New York, 1997).

\bibitem{heiselberg93:a}
H. Heiselberg, C. J. Pethick, and E. F. Staubo, Phys.\ Rev.\ Lett.\ {\bf 70}
  (1993) 1355.

\bibitem{glen95:c}
N. K. Glendenning and S. Pei, Phys.\ Rev. C {\bf 52} (1995) 2250.

\bibitem{glen97:c}
M. B. Christiansen and N. K. Glendenning, Phys.\ Rev.\ C {\bf 56} (1997) 2858.

\bibitem{glen97:a}
N. K. Glendenning, S. Pei and F. Weber, Phys.\ Rev.\ Lett.\ {\bf 79} (1997)
  1603.

\end{thebibliography}
\end{document}